\documentclass{iau} 
\usepackage{graphicx}

\title[The analogy of K-correction in the topic of gamma-ray bursts] 
{The analogy of K-correction in the topic of gamma-ray bursts}

\author[Levente Borv\'ak, Attila M\'esz\'aros \& Jakub \v{R}\'{\i}pa]   
{Levente Borv\'ak$^1$, Attila M\'esz\'aros$^2$ \& Jakub \v{R}\'{\i}pa$^{2,3,4}$}
\affiliation{$^1$Department of Physics, University of Dallas, 1845 East Northgate Drive, 
Irving, Texas, 75062 USA\\email:{\tt lborvak@udallas.edu}\\[\affilskip]
$^2$Astronomical Institute, Faculty of Mathematics and Physics, 
Charles University, CZ-180 00 Prague 8, V Hole\v{s}ovi\v{c}k\'ach 2, Czech Republic\\ email:{\tt meszaros@cesnet.cz}\\[\affilskip]
$^3$MTA-E\"otv\"os L\'or\'and University, Lend\"ulet Hot Universe Research Group, P\'azm\'any P\'eter s\'et\'any 1/A, Budapest, 1117, Hungary
\\$^4$E\"otv\"os L\'or\'and University, Institute of Physics, P\'azm\'any P\'eter s\'et\'any 1/A, Budapest, 1117, Hungary\\email:{\tt jripa@caesar.elte.hu}}

\pubyear{2018}
\volume{346}
\jname{High-mass X-ray binaries: illuminating the passage from
massive binaries to merging compact objects}
\editors{A.C. Editor, B.D. Editor \& C.E. Editor, eds.TO BE CHANGED}

\begin{document}

\maketitle

\begin{abstract}
It is well-known that there are two types of gamma-ray bursts (GRBs): short/hard and long/soft ones, respectively. The long GRBs are coupled to supernovae, but the short ones are associated with the so called macronovae (also known as kilonovae), which can serve as the sources of gravitational waves as well. The kilonovae can arise from the merging of two neutron-stars. The neutron stars can be substituded by more massive black holes as well. Hence, the topic of gamma-ray bursts (mainly the topic of short ones) and the topic of massive binaries, are strongly connected.

In this contribution, the redshifts of GRBs are studied. The surprising result - namely that the apparently fainter GRBs can be in average at smaller distances - is discussed again. In essence, the results of M\'esz\'aros et al. (2011) are studied again using newer samples of GRBs. The former result is confirmed by the  newer data.

\keywords{cosmology: miscellaneous, cosmology: observations,
gamma rays: bursts}
\end{abstract}

\firstsection 

\section{Inverse behavior of the peak-flux and fluence of GRBs:
The theory}

In the article \cite[M\'esz\'aros et al. (2011)]{Me11} (in what follows M11)
a remarkable property of the gamma-ray bursts (GRBs) was found. It can be briefly explained as follows (for details see the mentioned paper).

Given a GRB with measured peak-flux $P(z)$ (with the dimension of ph/(cm$^2$s), where ``ph'' means photon), if the object is at redshift $z$, then its isotropic peak-luminosity $\tilde{L}_\mathrm{iso}(z)$ (in units of ph/s) is related to the peak-flux by the expression
\begin{equation}
P(z) = \frac{(1+z) \tilde{L}_\mathrm{iso}(z)}{4\pi d_l(z)^2},
\end{equation}
where $d_l(z)$ is the luminosity distance of the object. 

An instrument measures the peak-flux at an interval $E_1 < E < E_2$, where $E_1$ and $E_2$ are 
the limiting photon energies given by the instrument, and $E$ is the measured energy of the photon. 
So the peak-luminosity must be taken from the interval $E_1(1 + z)$ and $E_2(1 + z)$, not simply from $E_1$ and $E_2$.

The same relation is also expected for the fluence if it has the dimension erg/cm$^2$.

It is standard cosmology that $d_l(z)$ increases with the redshift. For the exact formula, see \cite[Carroll et al. (1992)]{Ca92}. 
For any reasonable $\Omega$ factors it holds that $\lim_{z \rightarrow \infty} (d_l(z)/(1+z))$ is a finite value. 

However, because $\tilde{L}_\mathrm{iso}(z)$ depends on $z$, it is possible that $\tilde{L}_\mathrm{iso}(z)$ increases with $z$. In M11
it is argued that in some cases $\tilde{L}_\mathrm{iso}(z)$ can increase faster than $d_l(z)^2/(1+z)$ and 
hence an ``inverse'' behavior can occur: an apparently fainter GRB can be at a smaller redshift than a brighter one.

\section{Data of the {\em Swift} satellite in M11}

This theoretical expectation was shown to happen in M11 for {\em Swift}\footnote{https://swift.gsfc.nasa.gov} data. 
For the period from 20 November 2004 to 30 April 2010, 132 GRBs were observed by {\em Swift} with known redshifts. They were used in M11. 
We show here that this behavior also holds with the currently available Swift GRB database

\section{New sample from the {\em Swift} satellite} 

Below we show a comparison of the data used by M11 to the currently (until 10 May 2018) observed 416 GRBs. Five GRBs had no measured peak-fuxes, 4 GRBs had no measured fluences. Hence, 412 is the total number of GRBs with measured fluences (in the 15-350\,keV band) and $\tilde{E}_\mathrm{iso}$, but the total number of GRBs with measured peak-fuxes (1-s peak photon fluxes in the 15-350\,keV band) and $\tilde{L}_\mathrm{iso}$ is 411.
The total energy emitted by the GRBs, assuming isotropic emission, is denoted as 
$\tilde{E}_\mathrm{iso}$. The peak-luminosity, assuming isotropic emission, is denoted as 
$\tilde{L}_\mathrm{iso}$.

\begin{figure}[h]
\begin{center}
\includegraphics[width=5.2in]{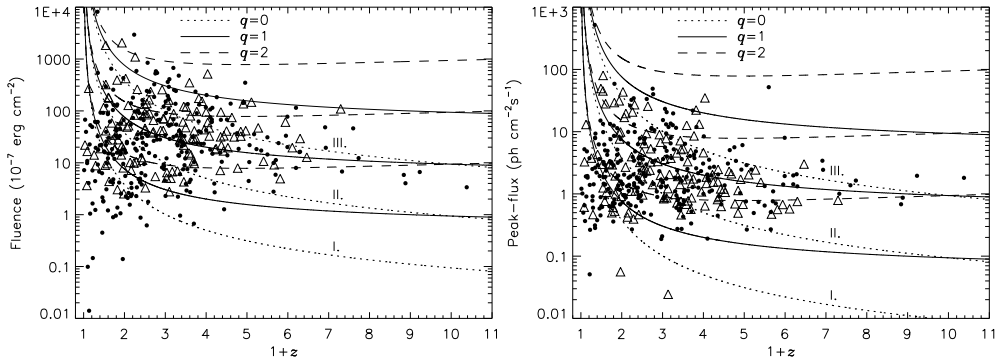}
 \caption{The 412 fluences (left panel) and the 411 peak-fluxes (right panel), respectively, versus $1+z$ are shown in the figure. The 132 GRBs mentioned in M11 are denoted by triangles. The curves with different $q$ illustrate the expected behavior, if either $\tilde{E}_\mathrm{iso}$ or                                                             
$\tilde{L}_\mathrm{iso}$ are proportional to $(1+z)^q$, i.e. $\tilde{E}_\mathrm{iso} = \tilde{E}_o(1+z)^q$ (constants $\tilde{E}_o$ are: I. $10^{51}$\,erg; II. $10^{52}$\,erg; III. $10^{53}$\,erg) and $\tilde{L}_\mathrm{iso} = \tilde{L}_o(1+z)^q$ (constants $\tilde{L}_o$ are: I. $10^{57}$\,ph/s; II. $10^{58}$\,ph/s; III. $10^{59}$\,ph/s). For the inverse behavior, a value of $q>1$ is needed for the biggest $z$. This trend was first mentioned by M11, and the larger sample indicate this possible behaviour, too. The figure is in essence an analogy of Fig. 3 of M11 with a larger sample.}
   \label{fig1}
\end{center}
\end{figure}

In Figure~\ref{fig1} these fluences and peak-fluxes, respectively, are compared with $1+z$. 

\begin{figure}[h]
\begin{center}
\includegraphics[width=2.6in]{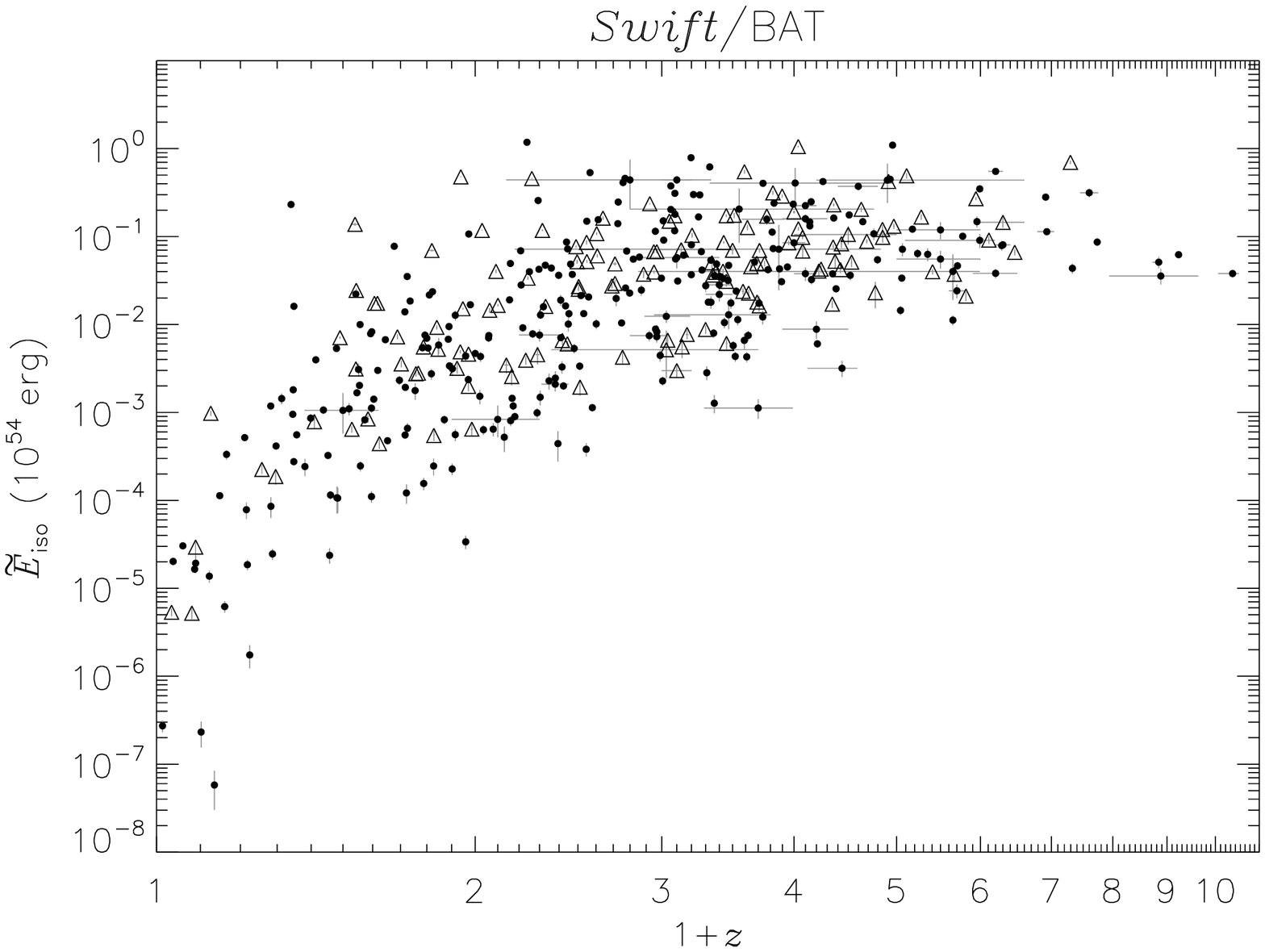}    
\includegraphics[width=2.6in]{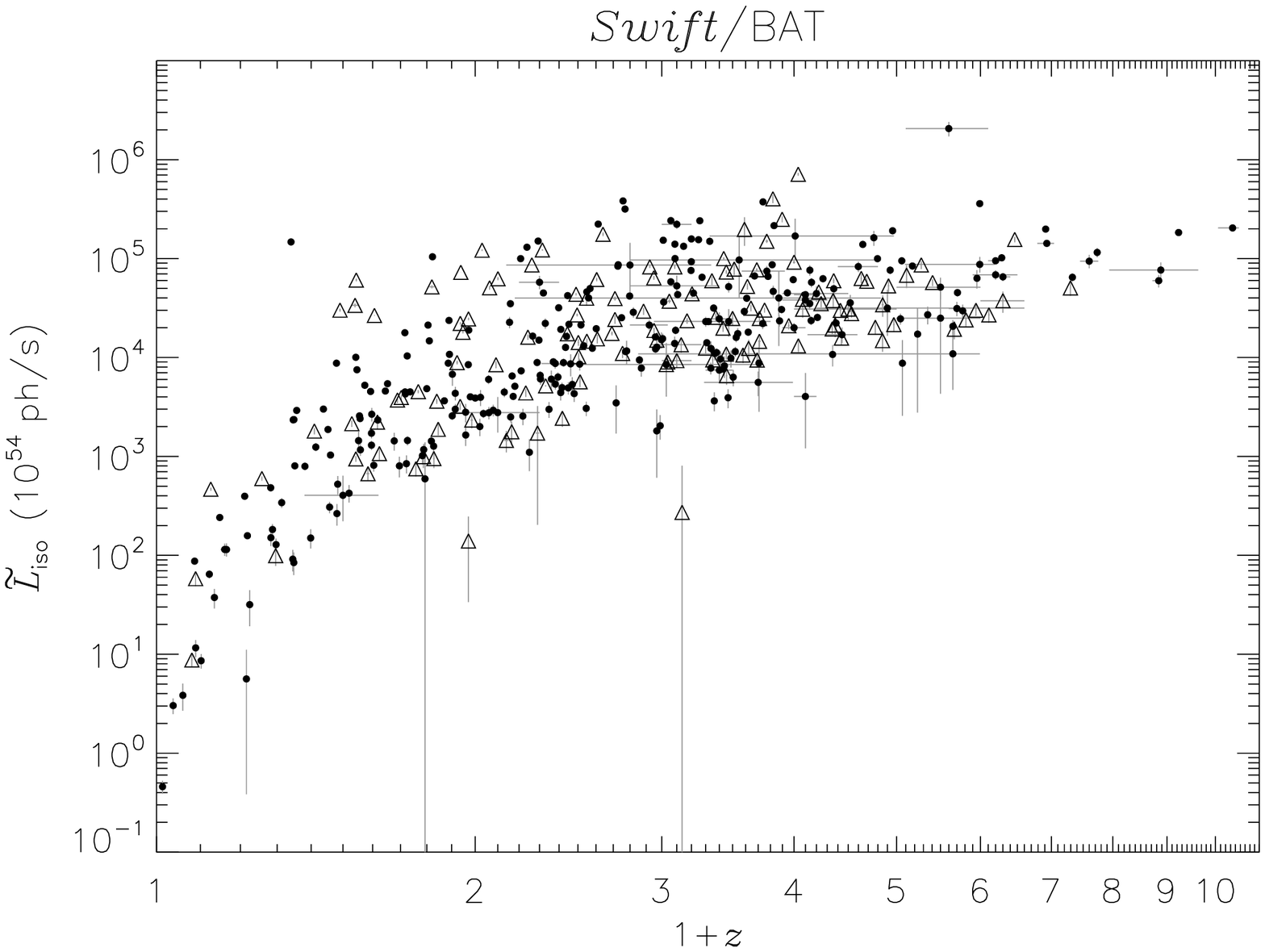}
 \caption{ $\tilde{E}_\mathrm{iso}$ and $\tilde{L}_\mathrm{iso}$, respectively, versus $1+z$ are shown in the figure. The number of all GRBs is 412 (411) on the left (right) plot. 
The 132 GRBs, mentioned in M11, are denoted by triangles. BAT means the Burst Alert Telescope. 
The figure is in essence an analogy of Fig. 4 of M11 with a larger sample.}
   \label{fig2}
\end{center}
\end{figure}

For the 412 fluences and 411 peak-fluxes, respectively, $\tilde{E}_\mathrm{iso}$ and
$\tilde{L}_\mathrm{iso}$ were calculated using Eq.(1.1).
These values are shown in Figure~\ref{fig2}.

\section{Remarks}

1. As a comparison for $\tilde{E}_\mathrm{iso}$, remember that $M_\odot c^2=1.8\times10^{54}$ erg, where $c$ is the velocity of light in vacuum, 
and $M_\odot$ is the mass of Sun.

2. If the photon has energy of $100y$ keV ($y$ is a positive unitless parameter),
then (because 100 keV=$1.6\times10^{-7}$ erg), $10^{58}$ ph/s corresponds 
to $1.6y\times10^{51}$ erg/s.

3. This behavior of $\tilde{E}_\mathrm{iso}$ and $\tilde{L}_\mathrm{iso}$ is analogous to the K-correction in optical astronomy 
(\cite[Hubble 1936]{Hub36}, \cite[Humason et al. 1956]{Hum56}).

\section{Conclusion}

The expanded dataset from the {\em Swift} satellite hints towards a possible inverse behavior that was first mentioned by M11.

\end{document}